\documentclass[prb,twocolumn,showpacs,preprintnumbers,amsmath]{revtex4}
\usepackage{dcolumn}
\usepackage{bm}
\usepackage{graphicx,color}

\newcommand{\ha}{$H_{\rm{a}}$}

\newcommand{\hco}{$H_{\rm{c1}}$}
\newcommand{\cv}{$\chi_{\rm{vol}}$}
\newcommand{\jc}{$J_{\rm{C}}$}
\newcommand{\tc}{$T_{\rm{C}}$}

\newcommand{\ti}{$T_{\rm{i}}$}

\newcommand{\figref}[1]{Fig.\,\protect\ref{#1}}
\newcommand{\cax}{Ca(Fe$_{1-x}$Co$_x$)$_2$As$_2$}
\newcommand{\bax}{Ba(Fe$_{1-x}$Co$_x$)$_2$As$_2$}
\newcommand{\ban}{Ba(Fe$_{0.93}$Co$_{0.07}$)$_2$As$_2$}
\newcommand{\cafs}{Ca(Fe$_{0.944}$Co$_{0.056}$)$_2$As$_2$}

\begin{document}
\title{Critical current and vortex dynamics in single crystals of Ca(Fe$_{1-x}$Co$_{x}$)$_2$As$_2$}

\author{A. K.~Pramanik}\affiliation{Institute for Solid State Research, IFW Dresden, D-01171 Dresden, Germany}
\author{L.~Harnagea}\affiliation{Institute for Solid State Research, IFW Dresden, D-01171 Dresden, Germany}
\author{S.~Singh}\affiliation{Institute for Solid State Research, IFW Dresden, D-01171 Dresden, Germany}
\author{S.~Aswartham}\affiliation{Institute for Solid State Research, IFW Dresden, D-01171 Dresden, Germany}
\author{G.~Behr}\affiliation{Institute for Solid State Research, IFW Dresden, D-01171 Dresden, Germany}
\author{S.~Wurmehl}\affiliation{Institute for Solid State Research, IFW Dresden, D-01171 Dresden, Germany}
\author{C.~Hess}\affiliation{Institute for Solid State Research, IFW Dresden, D-01171 Dresden, Germany}
\author{R.~Klingeler}\affiliation{Institute for Solid State Research, IFW Dresden, D-01171 Dresden, Germany}
\author{B.~B\"{u}chner}\affiliation{Institute for Solid State Research, IFW Dresden, D-01171 Dresden, Germany}

\date{\today}

\begin{abstract}
We investigate the critical current density and vortex dynamics in single crystals of
Ca(Fe$_{1-x}$Co$_{x}$)$_2$As$_2$ ($x$ = 0.051, 0.056, 0.065, and 0.073). The samples exhibit
different critical temperatures and superconducting phase fractions. We show that in contrast
to their Ba-based counterpart, the crystals do not exhibit a second peak in the field
dependence of magnetization. The calculated composition-dependent critical current density (\jc )
increases initially with Co doping, maximizing at $x$ = 0.065, and then decreases. This variation
in \jc\ follows the superconducting phase fractions in this series. The calculated \jc\ shows
strong temperature dependence, decreasing rapidly upon heating. Magnetic relaxation
measurements imply a nonlogarithmic dependence on time. We find that the relaxation rate is
large, reflecting weak characteristic pinning energy. The analysis of temperature- and field-
dependent magnetic relaxation data suggests that vortex dynamics in these compounds is consistent
with plastic creeping rather than the collective creep model, unlike other 122 pnictide
superconductors. This difference may cause the absence of the second peak in the field dependent
magnetization of Ca(Fe$_{1-x}$Co$_{x}$)$_2$As$_2$.
\end{abstract}

 \maketitle

\section {Introduction}

The recent discovery of superconductivity (SC) in Fe-based pnictides with chemical formula
LaO$_{1-x}$F$_x$FeAs (Ref.~\onlinecite{kamihara}) has attracted a lot of research activities in
the field of condensed matter physics in general and superconductivity in
particular.~\cite{takahashi,zhao,luetkens,christ,grafe,klauss,qazi,singh} Similar to cuprates,
pnictide superconductors also exhibit high critical temperature (\tc ) and type-II nature.
Moreover, SC arises from the layers, i.e., Fe-As layers in this case. In type-II superconductors,
magnetic fields above the lower critical field (\hco ) penetrate the bulk of the superconductor in the
form of flux lines or vortices, which in presence of an external current and/or magnetic field move, thus
causing finite dissipation to the transport current. This motion of vortices is further assisted by
thermal fluctuations. However, vortex movement can be hampered due to the pinning
barriers arising from disorders present in system. The measurement of isothermal magnetization ($M$) vs. field ($H$) and the magnetic relaxation are the most extensively used tools to study the vortex dynamics in a variety of superconducting
materials.~\cite{yeshurun,blatter} Commonly, the appearance of a second peak (SP)
in field dependent magnetization loop is believed to be directly associated with the
nature of pinning and thereof depending vortex creep mechanism.~\cite{abulafia,giller} Therefore,
understanding the vortex dynamics is one of the central issue in high-\tc\ superconductors
pertaining to both basic science as well as technological applications. In the case of the
cuprates, high anisotropy and small coherence lengths ($\xi$) render the pinning energy weak which in combination with their high working temperature results in giant flux creep and large magnetic
relaxation.~\cite{yeshurun,blatter} Although Fe-based superconductors share many aspects with the
cuprates but they have less anisotropy and larger $\xi$.~\cite{can-aniso} Therefore, it
remains a matter of interest to understand the vortex dynamics in these materials.

Here, we study the vortex dynamics by means of isothermal $M$($H$) as well as magnetic relaxation
measurements on single crystals of Ca(Fe$_{1-x}$Co$_{x}$)$_2$As$_2$ ($x$ = 0.051, 0.056, 0.065,
and 0.073). All samples are superconductors having varying \tc\ and superconducting phase
fractions. The parent compound CaFe$_2$As$_2$ belongs to the 122-family of Fe-based pnictides
which has the generic formula AFe$_2$As$_2$ where A is a alkaline earth element (i.e., Ba, Sr,
Ca). Unlike its 1111 analogue, 122 compounds are oxygen free and easier to grow with substantial
crystal size.~\cite{ni,can-Ca} At room temperature, all 122 compounds crystallize in the
ThCr$_2$Si$_2$-type tetragonal structure, however, the significant mismatch in ionic radii of
atoms at the A-site will presumably create local distortion, which will vary in different
systems. Around $T \simeq 170$~K, CaFe$_2$As$_2$ exhibits a first order structural phase
transition from a high-temperature tetragonal to a low-temperature orthorhombic symmetry which is
accompanied by formation of a spin density wave (SDW)-type antiferromagnetic (AFM)
order.~\cite{ronning,can-Ca} Partial doping with 3d elements like Co and Ni at the Fe-site or with
Na and K at the A-site yields electron or hole doping, respectively, in the system. Both the doping
suppresses the structural and SDW-transitions, and induces
SC.~\cite{rotter,sasmal,dong,neeraj1,neeraj2,rudi-Ca}

The vortex dynamics in Ba-122 superconducting compounds has been investigated extensively and
described within the framework of a collective pinning scenario. These studies yield the presence of a
SP in $M$($H$), a moderate critical current density \jc\ and strong vortex
pinning.~\cite{prozo-Ba07,yang,prozo-Ba,nakajima,kim,sun,shen,eskildsen} Moreover, the general
appearance of the SP in \bax\ remains irrespective of the crystal growth conditions. Interestingly, the
SP manifestation is less pronounced in strongly under- and over-doped
\bax\ where pinning is relatively weak as compared to the optical doping.~\cite{shen} In order to examine whether the presence
of the SP is common to the whole 122-family, we have studied the vortex dynamics in Co doped
Ca-122 compounds which exhibit similar \tc\ as their Ba-122 counterparts but differ in terms of
ionic radii of Ca$^{2+}$ and Ba$^{2+}$.~\cite{ni,rudi-Ca} To the best of our knowledge, the
present study is the first of its kind in Ca-122 compounds. Our study shows the absence of the SP in
$M$($H$) and we calculate relatively low critical currents \jc . Both findings are in contrast to
previous results on Ba-122. While our data imply a nonlogarithmic time dependence of the magnetic
relaxation, however, the relaxation rate is large indicating weak pinning potential. From the
detailed analysis of relaxation data we infer that vortex movement in Ca-122 occurs through the
plastic creeping which may explain the absence of the SP in Ca(Fe$_{1-x}$Co$_{x}$)$_2$As$_2$.

\section{Experimental Details}

Single crystals of Ca(Fe$_{1-x}$Co$_{x}$)$_2$As$_2$ with $x$ = 0.051, 0.056, 0.065, and 0.073 have
been grown from Sn-flux. The details of sample preparation and characterization will be presented
in elsewhere.~\cite{surjeet} Note, that in the present work the same crystals as in
Ref.~\onlinecite{rudi-Ca} have been used. The crystal structure was investigated by means of
powder x-ray diffraction (XRD) which implies impurity concentrations of less than 3\%. The
compositions have been determined by means of energy dispersive x-ray spectroscopy (EDX).
The EDX analysis is performed at different places of each sample. The variation in Co distribution is found within the limit of $x$ $\pm$ 0.003. The above mentioned Co concentration ($x$) is the average value obtained after several individual measurements. For the studies at hand, crystals of rectangular shape have been used with typical dimensions of (1-2)
$\times$ (1-2) $\times$ (0.3-0.4)~mm$^3$. Magnetization measurements have been performed in a
Quantum Design VSM-SQUID. Magnetic isotherms have been measured after cooling the samples in zero
magnetic field to the specific temperatures. Before collecting each $M$($H$) plot, the samples
have been heated much above their respective \tc 's and care has been taken for proper thermal
stabilization of the samples before measuring the isotherms. The typical field sweep rate for
$M$($H$) studies was 80~Oe/s. For magnetic relaxation measurements, the samples were zero field
cooled from well above $T_C$ to the selected temperatures ($T_i$) and after proper thermal stabilization the desired field ($H_a$) has been applied and magnetization has been measured as a function of time ($t$) for about 12000 s.
Similar to the $M$ vs. $H$ measurements, the samples have been warmed considerably above \tc\
before collecting $M$($t$).

\section{Results and discussions}

\begin{figure}[tb]
\includegraphics [angle=0,width=0.95\columnwidth,clip] {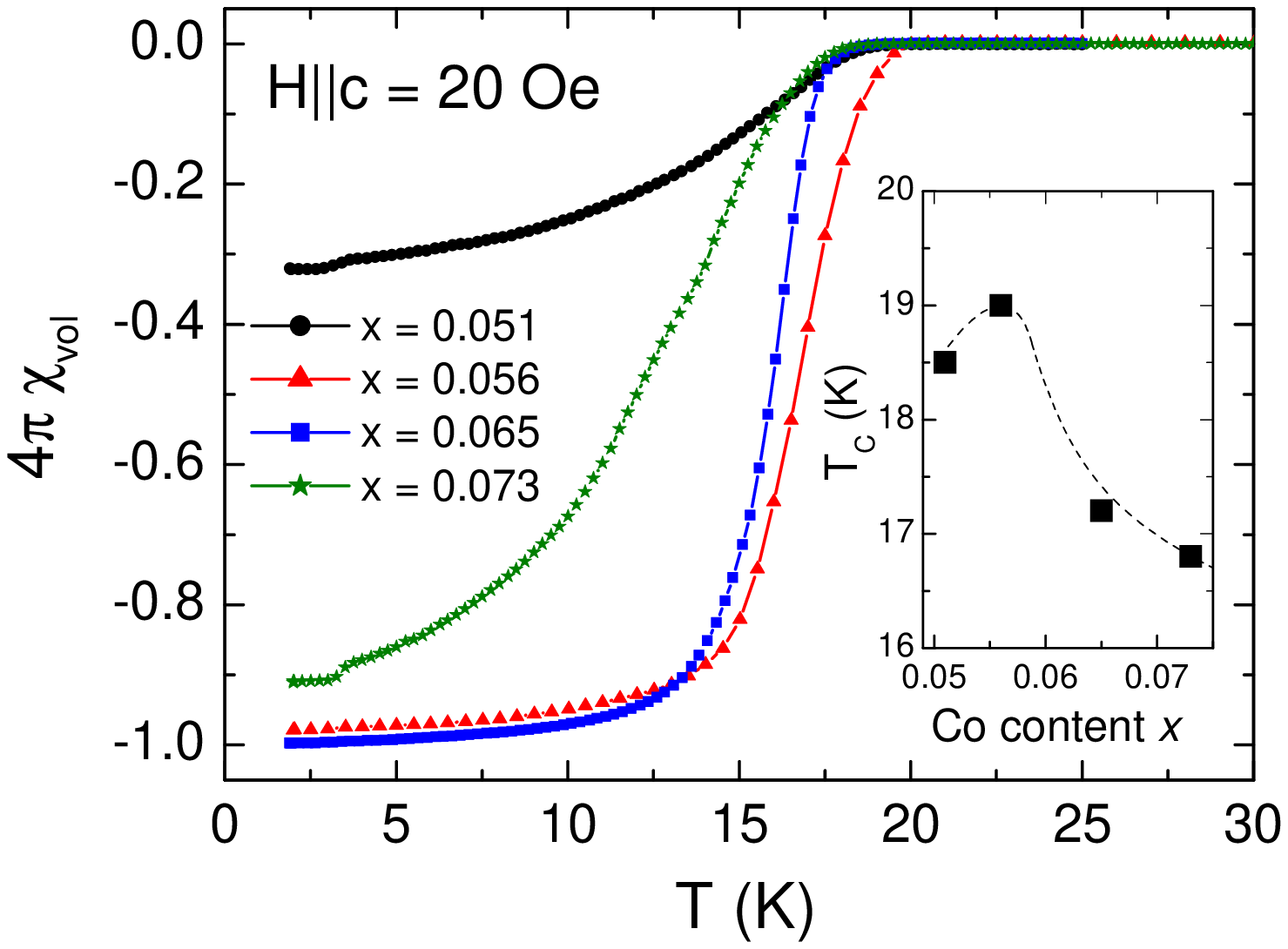}
\caption{(Color online) Volume susceptibility \cv\ of Ca(Fe$_{1-x}$Co$_{x}$)$_2$As$_2$ deduced
from dc magnetization measured in $H=20$~Oe with field parallel to the $c$-axis as a function of
temperature. The data have been obtained following zero field cooling protocol and were
corrected by the demagnetization factor. Inset: \tc\ vs. Co-content $x$. Line is guide to the eyes.}\label{fig:Fig1}
\end{figure}

Fig.~1 presents the temperature dependence of the volume susceptibility (\cv ) for
Ca(Fe$_{1-x}$Co$_{x}$)$_2$As$_2$ with $x$ = 0.051, 0.056, 0.065 and 0.073, respectively. \cv\ has
been deduced from dc-magnetization data measured at $H = 20$~Oe following the zero field cooling
(ZFC) protocol. The field has been applied parallel to the crystallographic $c$-axis. The data
have been corrected for the demagnetization effect where the demagnetization factor has been
estimated from the physical dimensions of the samples.~\cite{osborn} \figref{fig:Fig1} clearly
shows SC for all samples at low temperature. Upon progressively increasing the Co-content $x$, the
superconducting transition becomes sharper till $x$ = 0.065, and then it again broadens. The lower susceptibility and broad transition at $x$ = 0.051 and 0.073 suggest that a larger amount of inhomogeneity is associated
with these samples. This inhomogeneity is intrinsic to the samples as we do not find other chemically inhomogeneous phases within the resolution limit of XRD. This inhomogeneity may be associated with the coexistence of superconducting and SDW phases arising from the spontaneous electronic phase separation.~\cite{chu,park} We mention that, our EDX results imply reasonable homogeneous distribution of Co in macroscopic scale, and thus render a chemical inhomogeneity an unlikely explanation. The current measurements are not sufficient to determine the nature of inhomogeneity in these samples. A small kink in \cv ($T$) at low temperature around 3.5 K is most
probably associated with a small fraction of included Sn and is common to similar samples made
from Sn-flux.~\cite{ronning} We note that the superconducting phase fractions inferred from
\figref{fig:Fig1} are very similar to what has been reported for
Ba(Fe$_{1-x}$Co$_{x}$)$_2$As$_2$.~\cite{ni} The superconducting transition temperatures amount to
18.5, 19.0, 17.2, and 16.8~K for $x$ = 0.051, 0.056, 0.065, and 0.073, respectively (see the inset
of Fig.~1). Though there is only a small change in \tc\ within the studied compositions, the
superconducting volume fraction exhibits a clear doping dependence with a maximum around
$x=0.065$.~\cite{surjeet}

\begin{figure}[tb]
\includegraphics [angle=0,width=0.95\columnwidth,clip] {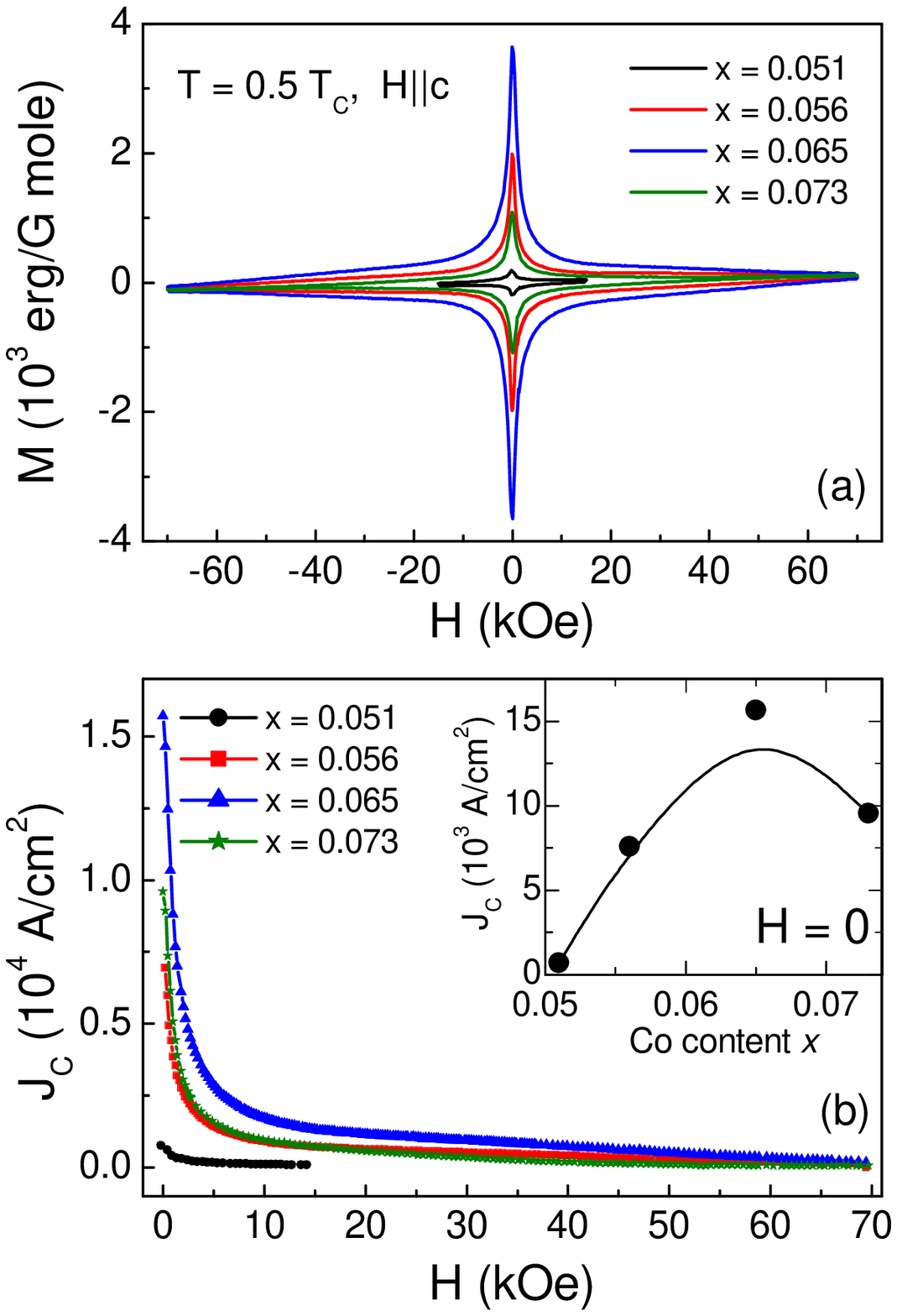}
\caption{(Color online) (a) The isothermal magnetization measured at temperature $\approx$
0.5T$_C$ as a function of field is plotted for Ca(Fe$_{1-x}$Co$_{x}$)$_2$As$_2$ series. The field
is applied parallel to crystallographic c axis. (b) The critical current density is plotted
against field for Ca(Fe$_{1-x}$Co$_{x}$)$_2$As$_2$ series with H $\parallel$ c axis. Inset shows
composition dependence of critical current density at H = 0 obtained from main panel. The line is a guide to the eyes.}
\label{fig:Fig2}
\end{figure}

Isothermal magnetization curves ($M$ vs. $H \parallel c$) collected at temperatures
$\approx$0.5T$_C$ are shown in \figref{fig:Fig2}. The data exhibit a central peak at zero magnetic
field and then magnetization decreases continuously with increasing magnetic fields. The sharp
peak around $H=0$ is similarly observed for other materials in the
122-family.~\cite{prozo-Ba07,kim,shen} Note, however, that the central peak appears to be
slightly sharper than found recently in \ban .~\cite{prozo-Ba07} Remarkably, there is no
SP in $M$($H$) within the measured field range for all samples which is in stark qualitative
contrast to previous results for \bax . Note, that a SP is also missing when measuring $M$ vs. $H$
with field perpendicular to the $c$-axis for $x$ = 0.056 (data not shown). Moreover, the $M$($H$) plots are rather symmetric with respect to the polarity of applied field.

From the irreversible part of the M vs H loop we have calculated the superconducting critical
current \jc\ exploiting the Bean critical state model which describes the penetration of
magnetic field into type-II superconductors:\cite{bean}

\begin{eqnarray}
    J_c = 20 \frac{\Delta M}{a \left(1-\frac{a}{3b} \right)}
\end{eqnarray}

where $\Delta M = M_{\rm dn} - M_{\rm up}$, $M_{\rm up}$ and $M_{\rm dn}$ is the magnetization
measured with increasing and decreasing field, respectively, $a$ and $b$ ($b > a$) are the
dimensions of the rectangular cross-section of the crystal normal to the applied
field.~\cite{footnote} The unit of $\Delta M$ is in emu/cm$^3$, $a$ and $b$ are in cm and the
calculated \jc\ is in A/cm$^2$. As expected from $M$($H$) in Fig.~2a, \jc\ also does not show a
SP, and decreases monotonically with field. Interestingly, it exhibits a pronounced doping
dependence but its values are more than one order of magnitude smaller than in the Ba-122
counterparts.~\cite{prozo-Ba07,yang,prozo-Ba,nakajima,kim,sun,shen} This low \jc\ indicates weak
pinning in \cax . The inset of Fig.~2b displays \jc\ at $H = 0$ for the different doping levels
$x$ under study. \jc\ initially increases with Co concentration showing a maximum at $x = 0.065$
and a decrease upon further increase of the Co content. Since \jc\ is associated with the pinning
of the vortices our data hence imply that pinning becomes stronger upon initial doping. In
particular, it appears that the variation of \jc ($x$) and thus of the pinning strength do not
follow the the variation of \tc ($x$) (see Fig. 1). The dome-like shape of \jc (x) in Fig.~2b is
similarly observed in superconducting \bax\ where it was shown that \jc\ enhances due to an
increase in intrinsic pinning arising from the domain walls of the coexisting AFM/orthorhombic
phase. These walls act as effective extended pinning centers which upon suppression of the related
ordering phenomena by doping become more fine and intertwined, thus giving rise to enhanced \jc\
when increasing $x$ in the underdoped regime.~\cite{prozo-Ba} Though our preliminary analysis
shows that the variation in \jc ($x$) follows the \tc ($x$) in Ca(Fe$_{1-x}$Co$_{x}$)$_2$As$_2$,
it needs to be further investigated whether such phase inhomogeneities play a crucial role in
controlling \jc\ in these samples.

\begin{figure}[tb]
\includegraphics [angle=0,width=0.95\columnwidth,clip] {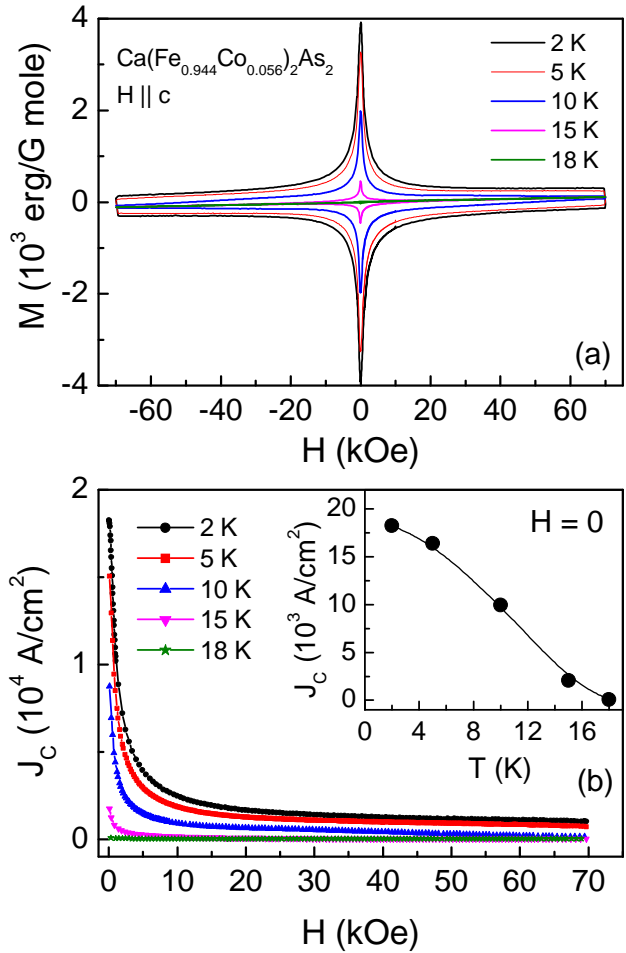}
\caption{(Color online) The field dependence of (a) magnetization and (b) critical current density
at different temperatures for \cafs . Inset: Temperature dependence of the critical current
density \jc\ at $H = 0$ for the same sample. The line is a guide to the eyes.} \label{fig:Fig3}
\end{figure}

In order to study the temperature variation of \jc\ we have measured the magnetization as a
function of field at different temperatures for \cafs\ which is the material with the highest
\tc\ in our study. Fig.~3a shows the recorded $M$($H$) curves at 2, 5, 10, 15 and 18~K, i.e.
within the superconducting regime. The width of the hysteresis increases with lowering the
temperature which suggests that the pinning characteristics becomes stronger at low temperature.
However, it is worth noting that we do not find the SP in $M$($H$) at any temperature below \tc .
In order to assess the pinning strength, we have again calculated \jc ($H$) following Eq.~1 from
the magnetization data (see Fig.~3b). With increasing temperature, \jc\ decreases as is can
already be inferred from the reduced $\Delta M$ in Fig. 3a. A strong temperature dependence of \jc
($H=0$) is demonstrated by the inset of Fig.~3b.

\begin{figure}[tb]
\includegraphics [angle=0,width=0.95\columnwidth,clip] {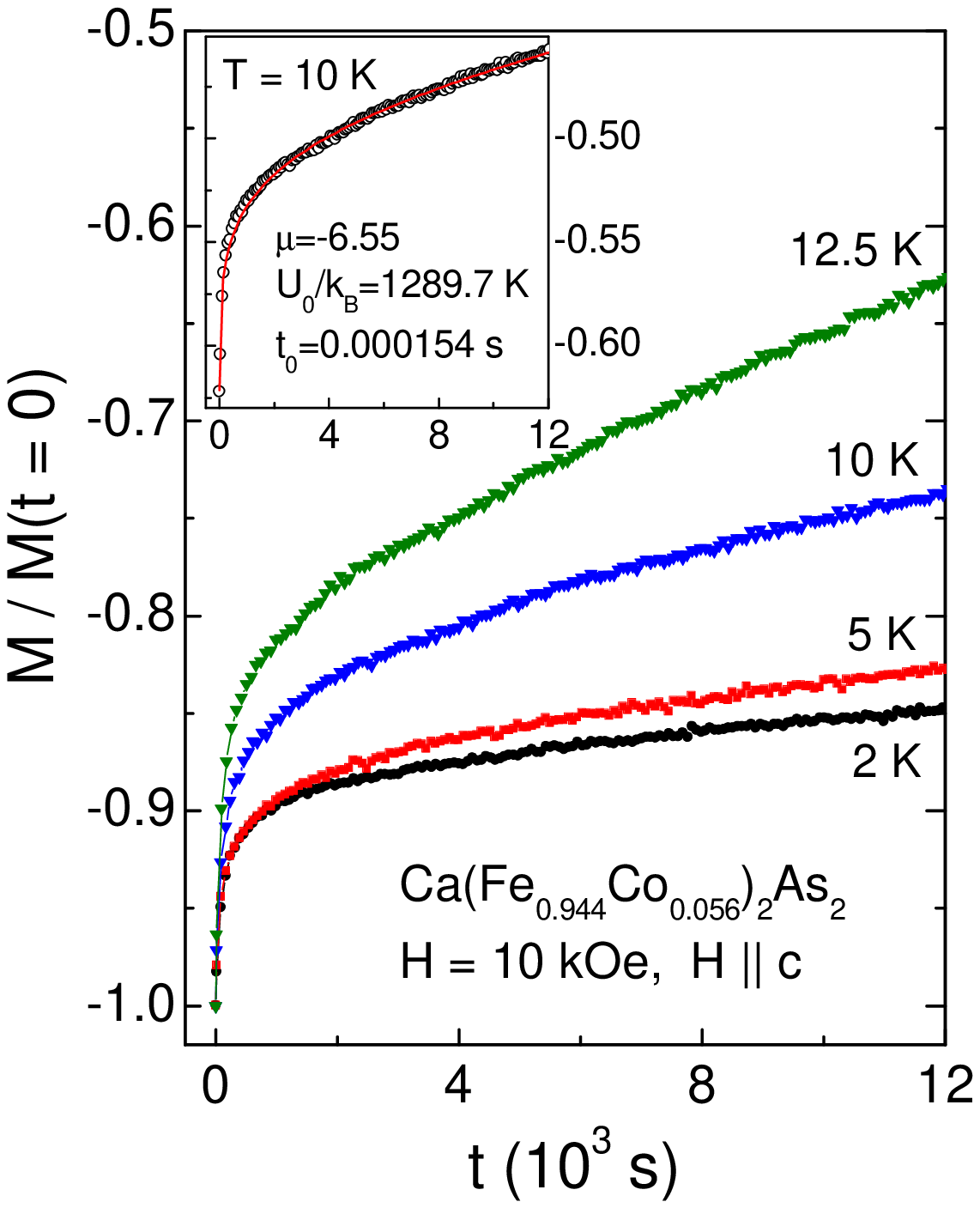}
\caption{(Color online) Time dependence of the ZFC magnetization as normalized at $t = 0$ recorded
at various temperatures. The magnetization is measured in $B$ = 10 kOe applied parallel to the $c$-axis
after cooling the sample in zero field to the desired temperature. In the inset, the best fit of
$M$($t$) at 10~K using Eq.~2 is shown with the obtained fitting parameters.}\label{fig:Fig4}
\end{figure}

Although the critical currents in \cax\ are relatively low, there is no SP effect visible. In
order to elucidate this observation, we have studied the vortex dynamics in further detail by
means of temperature- and field-dependent magnetic relaxation measurements in one of the
samples, i.e., \cafs . The results are displayed
in the main panel of Fig.~4. Here, $M$($t$) is shown at different temperatures which is normalized
to $M$ at $t=0$ in order to compare the relaxation. We have used \ha\ = 10~kOe (parallel to the
$c$-axis) and \ti\ = 2, 5, 10 and 12.5~K. In order to measure the field dependence, we fixed the
temperature at 10~K, and varied \ha , with \ha\ = 5, 7.5, 10, 15 and 20~kOe. The data in
\figref{fig:Fig4} show that the magnetization increases continuously with time in the complete
measurement period of 12000~s. For all temperatures, we observe significant increase of the
magnetization demonstrating high relaxation rates. For instance, at 10~K the magnetization
increases by almost 26\% within the measuring time span. However, it is evident in figure that
magnetization shows fast relaxation within initial time period of roughly 1000 s, and then relaxation
is relatively show. In addition, the figure shows that with
increasing temperature the relaxation rate becomes larger which indicates that the relaxation
process is thermally activated.

\begin{figure}[tb]
\includegraphics [angle=0,width=0.95\columnwidth,clip] {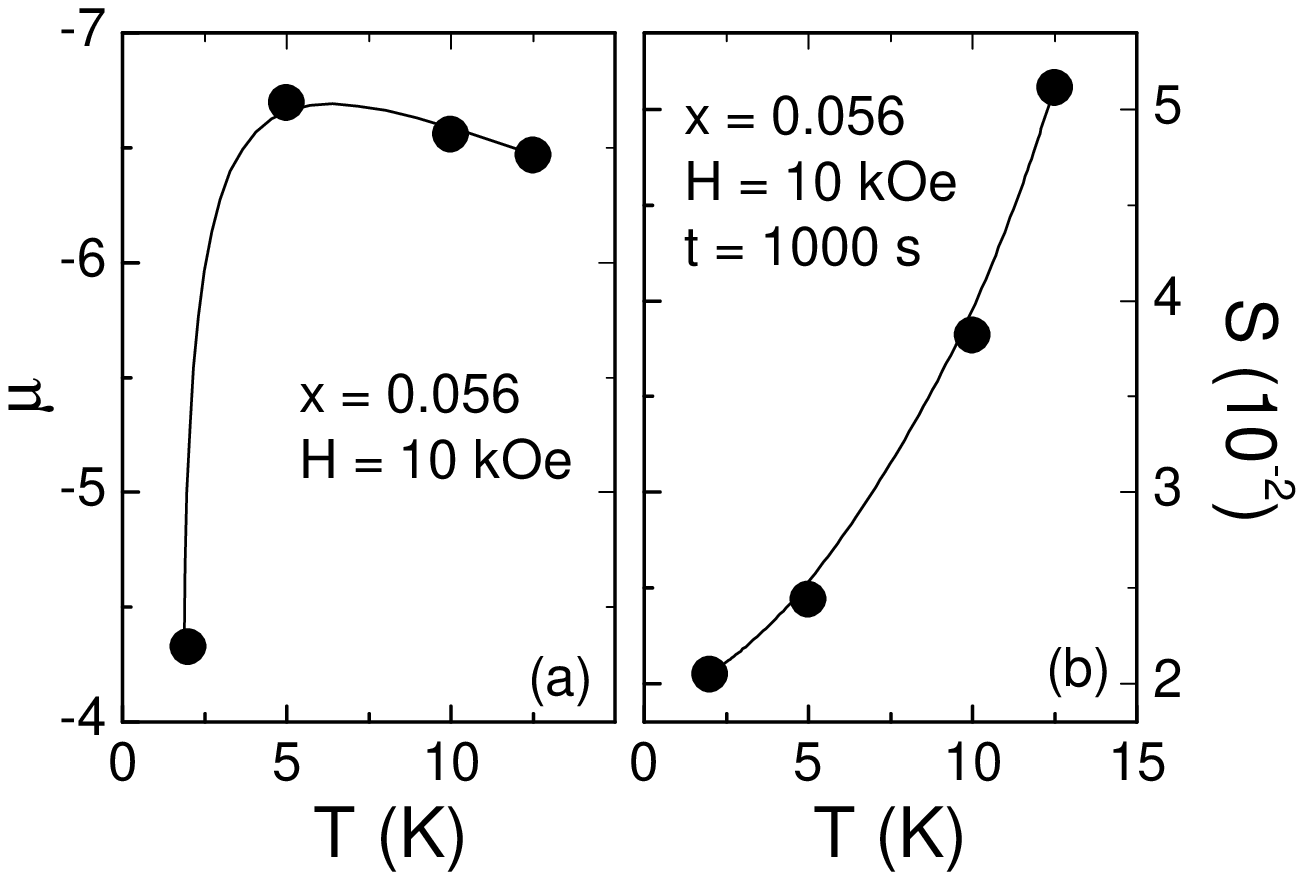}
\caption{(a) The parameter $\mu$ as obtained from fitting $M$($t$) by means of Eq.~2 as a function
of temperature. (b) The magnetic relaxation rate S calculated using Eq.~3 at $t = 1000$~s vs.
temperature (see the text). Lines are guides to the eyes.} \label{fig:Fig5}
\end{figure}

Magnetic relaxation in superconductors arises due to the nonequilibrium spatial distribution of
flux lines which are trapped at the pinning wells created by disorders. In the presence of an
externally applied magnetic field or current, flux lines experience a Lorentz force which drives the vortices
out of the pinning centers. In this scenario, \jc\ corresponds to the (critical) current density at
which the Lorentz force equals the maximum pinning force. However, this experimentally obtained supercurrent density,
$J$, is always lower than \jc\ due to the thermal fluctuations. The redistribution or creeping
of flux lines causes the change (or relaxation) of the magnetic moment with time. The magnetic
relaxation is basically determined by two competing factors: one is the localizing effect of
vortices arising from the pinning potential and another is the delocalizing effect realized by the
Lorentz force and the thermal depinning. The higher the pinning strength, the lower is the
relaxation of the magnetization. In the initial Anderson-Kim model~\cite{anderson} the barrier
energy $U$ is assumed to depend linearly on the current density as $U = U_0$(1 - $J/J_{\rm c}$),
which implies a logarithmic time dependence of the magnetization. However, more realistic
descriptions of energy barriers follow a nonlinear dependence on the current density. One of such
functional forms is the inverse power law barrier which has emerged from the collective vortex
pinning theory and is defined as $U = U_0$[($J_{\rm c}/J)^\mu - 1$]. This approach is commonly
known as `interpolation formula' according to which time dependence of the magnetization behaves
as:~\cite{feigel}

\begin{eqnarray}
    M(t) = M_0 \left[1 + \frac{\mu k_{\rm B}T}{U_0} \ln \left(\frac{t}{t_0}\right)\right]^{-1/\mu}
\end{eqnarray}

where $k_{\rm B}$ is the Boltzmann constant, $U_0$ is the barrier height in absence of a driving
force, $t_0$ is the characteristic relaxation time (10$^{-6}$ - 10$^{-12}$ s), and $\mu$ is the
field-temperature dependent parameter whose value is predicted to depend on the dimensionality of
the system as well as on the size of vortex bundles.~\cite{feigel} In 3 dimensional lattice, $\mu$ = 1/7, 3/2 and 7/9 for single vortex, small bundles and large bundles, respectively. Positive values of $\mu$ indicate a collective creep barrier while negative ones signal plastic creep.~\cite{prozo-Ba07,griessen} In the former case, the thermally activated creep occurs
through jumps of vortex bundles whereas in the latter one creep is driven by the sliding of vortex
lattice dislocations. From the Eq.~2, the normalized magnetization relaxation rate $S$
[=(1/$M$)d$M$/d$\ln$($t$)] can be deduced as following:~\cite{yeshurun}

\begin{eqnarray}
    S(t) = \frac{k_{\rm B}T}{U_0 + \mu k_{\rm B}T \ln(t/t_0)}
\end{eqnarray}

In order to exploit this model, we have measured magnetic relaxation at different temperatures and
fields. The measured magnetic relaxation data do not show a logarithmic dependence on time which
is expected in the pure Anderson-Kim model. However, we can describe the $M$($t$) data well in
terms of Eq.~2. One representative fit for data taken at $T=10$~K on \cafs\ is shown in the inset
of \figref{fig:Fig4}. Our analysis implies a negative value of the parameter $\mu$, thereby
indicating the plastic creep nature of vortices. Interestingly, the extracted value of $\mu$ is
significantly larger than predicted by theory~\cite{feigel}. However, higher values of $\mu$ than
the theoretical prediction have also been observed in both cuprate and pnictide
superconductors.~\cite{abulafia,prozo-Ba07} The high value of $\mu$ within the plastic creep
scenario signifies a high relaxation rate (Eq. 3) which is consistent with data in the Fig. 4.
This high $\mu$ and the initial fast increase of magnetic relaxation in Fig. 4 probably suggest
that in crystal there is a inhomogeneous distribution of pinning centers which render locally
correlated vortex bundles. As soon as magnetic field is applied, these bundles of flux lines slide
through the crystal giving rise to plastic creep. Once, however, this redistribution of vortex
bundles is settled down, the individual or bundle of vortex perform slow creeping, thus exhibiting
relatively slow relaxation. The temperature dependence of parameter $\mu$ (cf. \figref{fig:Fig5})
exhibits a significant increase of $|\mu|$ upon heating at low temperatures while for $T\gtrsim
5$~K there is only a very weak variation with a negative slope. The relaxation rate $S$ which is
calculated exploiting Eq.~3 is plotted as a function of temperature in Fig.~5b at $t = 1000$~s.
Even at low temperature, the calculated value of $S$ is much larger than in conventional
superconductors but comparable to cuprate and Ba-122
superconductors.~\cite{yeshurun,blatter,prozo-Ba07} With increasing temperature, $S$ exhibits a
significant monotonic increase instead of a plateau as it would be expected in collective pinning
theories of vortices.~\cite{yeshurun,blatter} The high value of $S$ is indicative of giant flux
motion which is in a qualitative agreement to the observed low \jc\ in our materials. On the other
hand, the positive slope d$S$/d$T$ is consistent with the scenario of plastic motion of flux
lines.~\cite{prozo-Ba07}

\begin{figure}[tb]
\includegraphics [angle=0,width=0.95\columnwidth,clip] {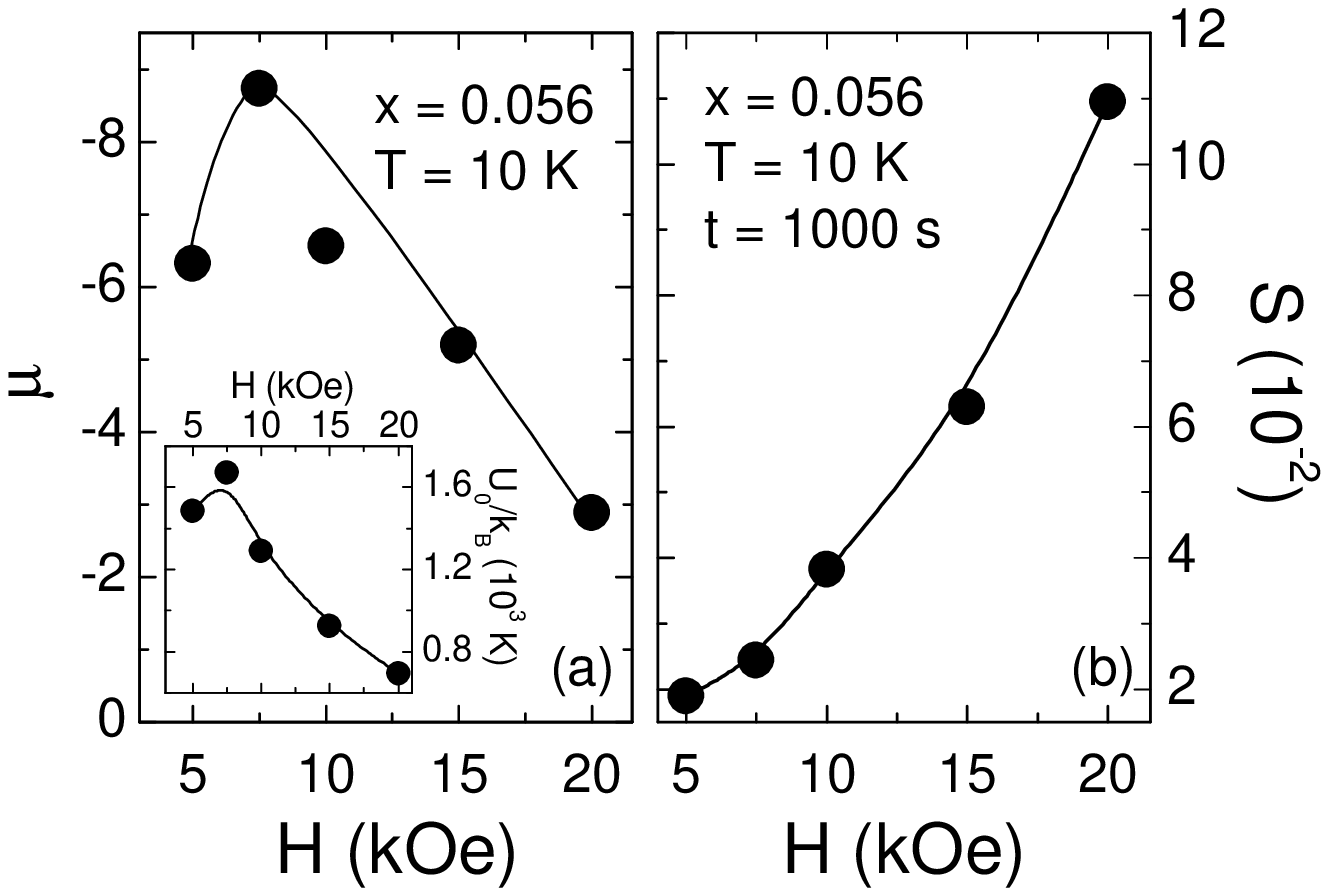}
\caption{Field dependence of (a) the parameter $\mu$ (see Eq.~2) and (b) the magnetic relaxation
rate at $t = 1000$~s (see Eq. 3) for \cafs . The inset of (a) displays the barrier energy $U_0$
obtained by means of Eq.~2 vs. magnetic field (see the text). Lines are guides to the eyes.}
\label{fig:Fig6}
\end{figure}

In order to further confirm the plastic nature of flux creeping we have measured $M$($t$) with varying
\ha\ at 10~K. Similar to Fig.~4, the data depict an increase of the magnetic relaxation rate for
higher \ha\ (not shown). Analysis of the data in terms of Eq.~2 again implies negative $\mu$
for all applied fields. As seen in \figref{fig:Fig6}a, there is a maximum in $|\mu |$ at $H_{\rm
m} \approx 7.5$~kOe. The field dependence of the barrier energy $U_0$ follows a very similar
behavior. Noteworthy, the decrease of $\mu$ and U$_0$ at $H_{\rm a} > H_{\rm m}$ contradicts the
collective creep model which implies $U_0 \propto H^\nu$, where $\nu$ is a positive
exponent.~\cite{abulafia} In contrast, it agrees well to the plastic creep model in which the
barrier energy depends on the field as $\propto$ 1/$\sqrt{H}$, thus decreasing with field. Further
information is provided by estimating the magnetic relaxation rate using Eq.~3 (Fig. 6b). $S$($H$)
exhibits a considerable monotonic increase which again corroborates a plastic nature of the creep
vortex dynamics in these compounds.~\cite{prozo-Ba07} Thus, the above experimental signatures
strongly suggest that the vortex dynamics in \cax\ is determined by the plastic creep of flux
lines.

The scenario of plastic vortex creep was invoked in cuprate superconductors due to failure of the
collective creep model in explaining the experimental observations in high fields above the
SP.~\cite{abulafia,giller} To be specific, it is argued that the SP in $M$($H$) is related to the
crossover from collective to plastic creeping of vortices. This crossover is attributed to
different magnetic field effects on the plastic and the collective energy barrier, respectively,
the former being smaller in high fields and thereby controlling the the vortex dynamics in this
field regime. Following this scenario, the absence of the SP in \cax\ can be explained by the weak
pinning energy and the absence of collective pinning, which is evident from our analysis above. To
be specific, weak pinning forces suggest that dislocations in the vortex lattice can easily move,
thereby causing dissipation and leading to plastic motion. However, we remind the contrasting
feature at low field in Fig.~6a, where both $\mu$(H) and U$_0$(H) increases. This observation
might suggest that the field associated with SP in these compounds is very low so that the
pronounced central peak just masks the SP. Note, that similar scenarios are discussed for very
underdoped and overdoped Ba(Fe$_{1-x}$Co$_{x}$)$_2$As$_2$ series where the weak pinning energy
leads to less pronounced SP in these samples.~\cite{shen}

The observed differences in vortex dynamics for \bax\ and \cax\ are quite intriguing. Both the
series differ in ionic radii of Ba$^{2+}$ and Ca$^{2+}$, otherwise they have comparable $T_C$ and
superconducting phase fractions. The Ca-122 compounds are relatively less anisotropic both in
terms of resistivity and upper critical field $H_{c2}$ which is expected to yield high pinning
energy in contrast to our observed behavior.~\cite{neeraj2,sun,tanatar-rho} However, during our
crystal growing studies, we observe that Ba-122 crystals can support As deficiency whereas Ca-122
materials do not allow such deficiency. This might create added pinning centers in Ba-122
crystals. It has been recently suggested for \bax\ that the structural domain walls in coexisting
AFM/orthorhombic phases create effective pinning centers. The presence of these structural domains
are common in 122 family.~\cite{tanatar-dom} With increasing $x$, it is shown that these walls
become finer and denser, and form interwoven pattern giving rise to stronger pinning. These are
favored by the suppression of orthorhombic distortion $\delta$ [= (a-b)/(a+b)] with increasing Co
concentration. For comparison, we have calculated $\delta$ for the parent compounds ($x$ = 0) with
values 0.36\% and 0.67\% for BaFe$_2$As$_2$ and CaFe$_2$As$_2$, respectively.~\cite{rotter1,krey}
With Co doping, this trend will presumably be retained. Due to higher degree of orthorhombic
distortion such domain walls will differ in Ca- and Ba-122 crystals, and this might explain the
different pinning in these compounds. Nonetheless, the effects of crystallographic parameters on
vortex pinning is quite significant for superconductors in general, and such pinning due to twin
boundaries is observed in YBa$_2$Cu$_3$O$_{7-y}$ (Ref. 43) and RNi$_2$B$_2$C (R = Er and Ho) (Ref.
44).

\section{Conclusion}

In summary, we have studied the vortex dynamics in single crystals of
Ca(Fe$_{1-x}$Co$_{x}$)$_2$As$_2$ ($x$ = 0.051, 0.056, 0.065 and 0.073) superconductors. Unlike
Ba-based 122 pnictides, we observe no SP in the field dependent magnetization loops. The critical
current density \jc\ calculated in the frame of Bean's model is low compared to other 122-family
members. With Co doping, \jc\ initially increases and then decreases which variation appears to be
consistent with that of the superconducting phase fractions. Moreover, the estimated \jc\ shows a
strong temperature dependence, decreasing rapidly upon heating. The magnetic relaxation exhibits a
nonlogarithmic time-dependence, however, the data are explained well by means of the interpolation
model. High magnetic relaxation rates imply weak pinning energy. Furthermore, from the analysis of
magnetic relaxation data we conclude that vortex dynamics in Co doped Ca-122 superconductors is
rather of plastic than collective nature. The comparison with other 122 superconductors might
indicate that the observed vortex creep behavior is connected with the strength of pinning energy
which varies with the A-site elements of 122 pnictides. However, to generalize this result similar
studies are needed in different samples including variety of doping species.\\

\begin{acknowledgments}
We acknowledge valuable discussions with V. Grinenko. We thank M. Deutschmann, S.
M\"uller-Litvanyi, R. M\"uller, J.~Werner, K.~Leger, and S.~Ga{\ss} for technical support. Work was
supported by the DFG through project BE 1749/13.
\end{acknowledgments}


\begin{thebibliography}{}
\bibitem[1]{kamihara} Y. Kamihara, T. Watanabe, M. Hirano, and H. Hosono, J. Am. Chem. Soc. \textbf{130}, 3296 (2008).
\bibitem[2]{takahashi} H. Takahashi, K. Igawa, K. Arii, Y. Kamihara, M. Hirano and H. Hosono, Nature \textbf{453}, 346 (2008).
\bibitem[3]{zhao} J. Zhao, Q. Huang, C. De La Cruz, S. Li, J. W. Lynn, Y. Chen, M. A. Green, G. F. Chen, G. Li, Z. Li, J. L. Luo, N. L. Wang and P. Dai, Nature Mater. \textbf{7}, 953 (2008).
\bibitem[4]{christ} A. D. Christianson, E. A. Goremychkin, R. Osborn, S. Rosenkranz, M. D. Lumsden, C. D. Malliakas, I. S. Todorov, H. Claus, D. Y. Chung, M. G. Kanatzidis, R. I. Bewley and T. Guidi, Nature \textbf{456}, 930 (2008).
\bibitem[5]{luetkens} H. Luetkens, H.-H. Klauss, M. Kraken, F. J. Litterst, T. Dellmann, R. Klingeler, C. Hess, R. Khasanov, A. Amato, C. Baines, M. Kosmala, O. J. Schumann, M. Braden, J. Hamann-Borrero, N. Leps, A. Kondrat, G. Behr, J.Werner and B. Buchner, Nature Mater. \textbf{8}, 305 (2009).
\bibitem[6]{grafe} H.-J. Grafe, D. Paar, G. Lang, N. J. Curro, G. Behr, J. Werner, J. Hamann-Borrero, C. Hess, N. Leps, R. Klingeler, and B. Buchner, Phys. Rev. Lett. \textbf{101}, 047003 (2008).
\bibitem[7]{klauss} H.-H. Klauss, H. Luetkens, R. Klingeler, C. Hess, F. J. Litterst, M. Kraken, M. M. Korshunov, I. Eremin,
S.-L. Drechsler, R. Khasanov, A. Amato, J. Hamann-Borrero, N. Leps, A. Kondrat, G. Behr, J. Werner, and B. Buchner, Phys. Rev. Lett. \textbf{101}, 077005 (2008).
\bibitem[8]{qazi} M. M. Qazilbash, J. J. Hamlin, R. E. Baumbach, Lijun Zhang, D. J. Singh, M. B. Maple and D. N. Basov, Nature Phys. \textbf{5}, 647 (2009).
\bibitem[9]{singh} D. J. Singh and M.-H. Du, Phys. Rev. Lett. \textbf{100}, 237003 (2008).
\bibitem[10]{yeshurun} Y. Yeshurun, A. P. Malozemoff and A. Shaulov, Rev. Mod. Phys. \textbf{68}, 911 (1996).
\bibitem[11]{blatter} G. Blatter, M. V. Feigelman, V. B. Geshkenbein, A. I. Larkin and V. M. Vinokur, Rev. Mod. Phys. \textbf{66}, 1125 (1994).
\bibitem[12]{abulafia} Y. Abulafia, A. Shaulov, Y. Wolfus, R. Prozorov, L. Burlachkov, and Y. Yeshurun, D. Majer and E. Zeldov, H. W{\"u}hl, V. B. Geshkenbein, and V. M. Vinokur, Phys. Rev. Lett. \textbf{77}, 1596 (1996).
\bibitem[13]{giller} D. Giller, A. Shaulov, R. Prozorov, Y. Abulafia, Y. Wolfus, L. Burlachkov, Y. Yeshurun, E. Zeldov, V. M. Vinokur, J. L. Peng and R. L. Greene, Phys. Rev. Lett. \textbf{79}, 2542 (1997).
\bibitem[14]{can-aniso} M. A. Tanatar, N. Ni, C. Martin, R. T. Gordon, H. Kim, V. G. Kogan, G. D. Samolyuk, S. L. Budko, P. C. Canfield, and R. Prozorov, Phys. Rev. B \textbf{79}, 094507 (2009).
\bibitem[15]{ni} N. Ni, M. E. Tillman, J.-Q. Yan, A. Kracher, S. T. Hannahs, S. L. Budko, and P. C. Canfield, Phys. Rev. B \textbf{78}, 214515 (2008).
\bibitem[16]{can-Ca} N. Ni, S. Nandi, A. Kreyssig, A. I. Goldman, E. D. Mun, S. L. Budko, and P. C. Canfield, Phys. Rev. B \textbf{78}, 014523 (2008).
\bibitem[17]{ronning} F. Ronning, T Klimczuk, E D Bauer, H Volz and J D Thompson, J. Phys.: Condens. Matter \textbf{20} 322201 (2008).
\bibitem[18]{rotter} M. Rotter, M. Tegel, and D. Johrendt, Phys. Rev. Lett. \textbf{101}, 107006 (2008).
\bibitem[19]{sasmal} K. Sasmal, B. Lv, B. Lorenz, A. M. Guloy, F. Chen, Yu-Yi Xue, and Ching-Wu Chu, Phys. Rev. Lett. \textbf{101}, 107007 (2008).
\bibitem[20]{dong} J K Dong, L Ding, H Wang, X F Wang, T Wu, G Wu, X H Chen and S Y Li, New J. Phys. \textbf{10}, 123031 (2008).
\bibitem[21]{neeraj1} Neeraj Kumar, Songxue Chi, Ying Chen, Kumari Gaurav Rana, A. K. Nigam, A. Thamizhavel, W. Ratcliff, S. K. Dhar, and J. W. Lynn, Phys. Rev. B \textbf{80}, 144524 (2009).
\bibitem[22]{neeraj2} Neeraj Kumar, R. Nagalakshmi, R. Kulkarni, P. L. Paulose, A. K. Nigam, S. K. Dhar, and A. Thamizhavel, Phys. Rev. B \textbf{79}, 012504 (2009).
\bibitem[23]{rudi-Ca} R. Klingeler, N. Leps, I. Hellmann, A. Popa, U. Stockert, C. Hess, V. Kataev, H.-J. Grafe, F. Hammerath, G. Lang, S. Wurmehl, G. Behr, L. Harnagea, S. Singh, and B. B{\"u}chner, Phys. Rev. B \textbf{81}, 024506 (2010).
\bibitem[24]{prozo-Ba07} R. Prozorov, N. Ni, M. A. Tanatar, V. G. Kogan, R. T. Gordon, C. Martin, E. C. Blomberg, P. Prommapan, J. Q. Yan, S. L. Budko, and P. C. Canfield, Phys. Rev. B \textbf{78}, 224506 (2008).
\bibitem[25]{yang} Huan Yang, Huiqian Luo, Zhaosheng Wang, and Hai-Hu Wena, Appl. Phys. Lett. \textbf{93}, 142506 (2008).
\bibitem[26]{prozo-Ba} R. Prozorov, M. A. Tanatar, N. Ni, A. Kreyssig, S. Nandi, S. L. Budko, A. I. Goldman, and P. C. Canfield, Phys. Rev. B \textbf{80}, 174517 (2009).
\bibitem[27]{nakajima} Y. Nakajima, Y. Tsuchiya, T. Taen, T. Tamegai, S. Okayasu, and M. Sasase, Phys. Rev. B \textbf{80}, 012510 (2009).
\bibitem[28]{kim} Hyeong-Jin Kim, Yong Liu, Yoon Seok Oh, Seunghyun Khim, Ingyu Kim, G. R. Stewart, and Kee Hoon Kim, Phys. Rev. B \textbf{79}, 014514 (2009).
\bibitem[29]{sun} D. L. Sun, Y. Liu, and C. T. Lin, Phys. Rev. B \textbf{80}, 144515 (2009).
\bibitem[30]{shen} Bing Shen, Peng Cheng, Zhaosheng Wang, Lei Fang, Cong Ren, Lei Shan, and Hai-Hu Wen, Phys. Rev. B \textbf{81}, 014503 (2010).
\bibitem[31]{eskildsen} M. R. Eskildsen, L. Ya. Vinnikov, T. D. Blasius, I. S. Veshchunov, T. M. Artemova, J. M. Densmore, C. D. Dewhurst, N. Ni, A. Kreyssig, S. L. Budko, P. C. Canfield, and A. I. Goldman, Phys. Rev. B \textbf{79}, 100501(R) (2009).
\bibitem[32]{surjeet} S. Singh, L. Harnagea, S. Wurmehl, R. Klingeler, C. Hess, G. Behr, and B. B{\"u}chner (unpublished).
\bibitem[33]{osborn} J. A. Osborn, Phys. Rev. \textbf{67}, 351 (1945).
\bibitem[34]{chu} Jiun-Haw Chu, James G. Analytis, Chris Kucharczyk, and Ian R. Fisher, Phys. Rev. B \textbf{79}, 014506 (2009).
\bibitem[35]{park} J. T. Park, D. S. Inosov, Ch. Niedermayer, G. L. Sun, D. Haug, N. B. Christensen, R. Dinnebier, A.V. Boris,
A. J. Drew, L. Schulz, T. Shapoval, U. Wolff, V. Neu, Xiaoping Yang, C. T. Lin, B. Keimer, and V. Hinkov, Phys. Rev. Lett. \textbf{102}, 117006 (2009).
\bibitem[36]{bean} C. P. Bean, Phys. Rev. Lett. \textbf{8}, 250 (1962); Rev. Mod. Phys. \textbf{36}, 31 (1964).
\bibitem[37]{footnote}The labeling of the crystal dimensions is in accordance to Fig.~1 in
Ref.~\onlinecite{prozo-Ba07}.
\bibitem[38]{anderson} P. W. Anderson and Y. B. Kim, Rev. Mod. Phys. \textbf{36}, 39 (1964).
\bibitem[39]{griessen} R. Griessen, A. F. T. Hoekstra, H. H. Wen, G Doornbos, and H. Schnack, Physica C \textbf{282-287}, 347 (1997).
\bibitem[40]{feigel} M. V. Feigel'man, V. B. Geshkenbein, A. I. Larkin, and V. M. Vinokur, Phys. Rev. Lett. \textbf{63}, 2303 (1989).
\bibitem[41]{tanatar-rho} M. A. Tanatar, N. Ni, G. D. Samolyuk, S. L. Budko, P. C. Canfield, and R. Prozorov, Phys. Rev. B \textbf{79}, 134528 (2009).
\bibitem[42]{tanatar-dom} M. A. Tanatar, A. Kreyssig, S. Nandi, N. Ni, S. L. Budko, P. C. Canfield, A. I. Goldman, and R. Prozorov, Phys Rev. B \textbf{79}, 180508(R) (2009).
\bibitem[43]{rotter1} Marianne Rotter, Marcus Tegel, and Dirk Johrendt, Inga Schellenberg, Wilfried Hermes, and Rainer P$\ddot{o}$ttgen, Phys. Rev. B \textbf{78}, 020503(R) (2008).
\bibitem[44]{krey} A. Kreyssig, M. A. Green, Y. Lee, G. D. Samolyuk, P. Zajdel, J. W. Lynn, S. L. Bud'ko,
M. S. Torikachvili, N. Ni, S. Nandi, J. B. Le$\tilde{a}$o, S. J. Poulton, D. N. Argyriou, B. N.
Harmon, R. J. McQueeney, P. C. Canfield, and A. I. Goldman, Phys. Rev. B \textbf{78}, 184517
(2008).
\bibitem[45]{vinnikov-YBCO} L. Ya. Vinnikov, L. A. Gurevich, G. A. Yemelchenko and Yu. A. Ossipyan, Solid State Commun. \textbf{67}, 421 (1988).
\bibitem[46]{vinnikov} L. Ya. Vinnikov, J. Anderegg, S. L. Bud'ko, P. C. Canfield, and V. G. Kogan, Phys. Rev. B \textbf{71}, 224513 (2005).

\end{thebibliography}
\end{document}